# The Necessity of Acceptance & Rejection in Inferential Reasoning

**Khashayar Irani**
**Birkbeck College, University of London**
**k.irani@mathematicallogic.com**
**November 2026**

**Abstract**

In this paper, we argue that acceptance and rejection constitute two fundamental positions necessary for inferential reasoning.[1] Our central thesis is that a proposition acquires an inferential role when it is considered from a position of acceptance or rejection, rather than purely understood as propositional content. We distinguish between understanding a proposition, entertaining it hypothetically, and employing it within an inference. Particular attention is given to hypothetical reasoning and suspended judgement, which appear to challenge the necessity of adopting either position.[2] We maintain that these cases do not undermine our thesis, since inferential reasoning can involve provisional acceptance or rejection without requiring personal belief or permanent commitment. Accordingly, the necessity of these positions should be understood as inferential rather than psychological. By examining the relationship between propositional content, inferential positions, and rational commitment, we defend the view that acceptance and rejection provide the fundamental conditions under which propositions can function within reasoning. This establishes a philosophical foundation for understanding inference through opposed rational positions.

[1] The present investigation deliberately adopts a non-technical approach to acceptance and rejection, seeking to clarify their philosophical necessity for inferential reasoning without introducing a formal logical system. Its primary objective is to explain why understanding propositional content alone is insufficient for its employment within inference and why acceptance and rejection constitute fundamental, distinguishable positions of rational assessment. This approach differs from the technical bilateral framework developed in Khashayar Irani (2026a), where acceptance and rejection are represented as positive and negative polar positions towards propositional content. In that formal framework, +A represents the acceptance of proposition A, whereas −A represents its rejection; the positive and negative signs function as structural indicators of the respective inferential positions rather than as logical constants. That investigation formally distinguishes rejection from propositional negation and examines their systematic relationship through negation, reflexivity, coherency and incoherency. The present enquiry also differs from Anna Gomolińska's (1998) which develops AEL2 as a nonmonotonic formalism for representing the knowledge states and decisions of an introspective agent, including the possibility of withholding judgement. Rather than constructing a bilateral calculus or an epistemic formalism, our intention here is to establish a philosophical account of the inferential significance and necessity of acceptance and rejection, accessible independently of technical logical machinery. Furthermore, by inferential reasoning, we mean the establishment of logical consequence through the application of inference rules governing logical constants, as exemplified by natural deduction systems. Within this framework, the inferential relationship between premises and conclusions is determined by the rules specifying the legitimate introduction and elimination of logical constants. We distinguish inferential reasoning in this sense from deductive reasoning understood semantically, where logical consequence is characterised by the preservation of truth from premises to conclusions. Therefore, the former concerns the rule-governed derivability of conclusions, whereas the latter concerns the truth-preserving relationship between premises and conclusions. For a comprehensive examination of the inferential roles of logical constants, see Irani (2025).

[2] Alongside the present manuscript, we are developing a further independent paper, entitled *Suspended Judgement: Epistemic Neutrality and the Limits of Inferential Commitment*, which examines the philosophical foundations and inferential significance of the concept of suspended judgement.

The capacity to understand propositions constitutes an essential feature of rational thought, yet understanding alone does not explain how propositions participate in inferential reasoning. When individuals encounter a statement, they may recognise its meaning, appreciate the circumstances it describes, and identify its possible implications without adopting any particular position towards its truth. Consider someone reading a newspaper report claiming that a proposed economic policy will reduce unemployment. The reader may understand the claim, recognise its relevance to public policy, and appreciate the reasoning offered in its support without accepting the prediction or rejecting its plausibility. Such comprehension establishes familiarity with propositional content but does not necessarily determine whether that content should function as a reason for further conclusions. A similar distinction arises when a student understands a philosophical argument without being persuaded by its premises. The student possesses sufficient conceptual understanding to explain the argument, although this ability does not establish an inferential commitment to the propositions involved. These examples suggest that understanding and reasoning, although closely connected, cannot be treated as equivalent intellectual activities. Understanding makes propositional content available for consideration, whereas employing that content within an inference requires determining the role it occupies in relation to other propositions. The philosophical question is therefore not simply how individuals comprehend statements, but what enables those statements to become operative within the practice of giving and evaluating reasons. We argue that acceptance and rejection constitute two fundamental positions through which propositions acquire such inferential significance. Acceptance allows a proposition to function positively within an argument, while rejection establishes a negative position from which its admissibility or implications may be challenged.[3] These positions should not be confused with psychological states such as confidence, conviction, or emotional attachment. Rather, they concern the rational treatment of propositions within particular contexts of reasoning. Their significance emerges from the distinction between possessing an understanding of what a statement expresses and determining how that statement may contribute to the justification, assessment, or criticism of further claims. Establishing this distinction provides the starting point for investigating whether acceptance and rejection are indispensable conditions of inferential reasoning, without presupposing that every proposition must immediately become the object of a settled judgement.

[3] One important meaning-theoretic application of acceptance and rejection is the formulation of a bilateral system of reasoning intended to accommodate logical inferentialism within a classical logical framework. For a detailed examination of logical inferentialism and the challenges it poses to classical logic, see Irani (2025).

The distinction between understanding and inferential employment becomes clearer when acceptance and rejection are examined independently. Acceptance involves adopting a positive position towards a proposition, permitting it to contribute to the development of an argument as a premise or supporting consideration.[4] Suppose a building inspector receives a technical report concerning the structural condition of a residential property. Understanding the report enables the inspector to identify its claims and appreciate their significance, but accepting particular findings allows those claims to support recommendations concerning necessary repairs. Acceptance does not guarantee that the findings are correct; rather, it establishes their operative status within the inspector's reasoning. Rejection performs a different function because it involves taking a negative position towards propositional content.[5] If subsequent examination reveals that a reported measurement was obtained using defective equipment, the inspector may reject the corresponding claim and exclude it from the assessment. Such rejection cannot be identified with purely failing to accept the measurement. Before examining the equipment, the inspector might have remained uncertain about its reliability without possessing grounds for opposing the claim. This difference reveals why rejection must be distinguished from non-acceptance. A person who has not accepted a proposition may be undecided, insufficiently informed, or unwilling to reach a judgement, whereas someone who rejects it adopts a position against its admissibility or correctness within the relevant inquiry. The distinction also prevents rejection from being reduced to a passive absence of positive commitment. Although acceptance and rejection may concern the same propositional content, they introduce different rational responsibilities and affect the development of reasoning in distinguishable ways. Acceptance permits certain conclusions to be supported by a proposition, while rejection may challenge an assumption, exclude an unreliable consideration, or provide grounds for reconsidering an argument. Neither position can be adequately explained purely as the absence of the other, since the failure to reject a proposition does not establish its acceptance any more than the failure to accept it establishes rejection. Their independence is therefore not a matter of linguistic preference but reflects the different functions through which propositions are treated within rational assessment. Recognising these functions allows us to investigate why inferential reasoning requires more than the comprehension of propositional meaning.

---

[4] For a classical treatment of bilateralism, in which assertion and denial are assigned distinct roles within logical reasoning, see Ian Rumfitt (2000).

[5] The distinction between rejection and propositional negation is both logically and philosophically significant. Rejection expresses a negative inferential position towards a proposition, whereas negation is a logical operator that operates upon propositional content. Accordingly, rejecting a proposition must be distinguished from accepting its negation, since the former concerns the polarity adopted towards the original proposition, while the latter involves accepting a proposition whose content has been negated. Yet, within a classical bilateral framework, these two distinct positions are systematically related through negation, such that rejection of a proposition corresponds inferentially to acceptance of its negation, and acceptance of a proposition corresponds to rejection of its negation. Thus, their classical inferential equivalence does not eliminate the fundamental distinction between negative polarity and negated propositional content. For a technical examination of these relations, see Irani (2026a).

Inferential reasoning is not purely the possession of meaningful statements or the recognition of relations between their contents; it is also a practice in which propositions are employed to support, examine, and challenge conclusions. This practical dimension may be described as a pragmatic inferential space, provided that the term 'pragmatic' is understood as referring to the employment of propositions within reasoning rather than to a comprehensive philosophical doctrine. Consider a local authority deciding whether to introduce restrictions on vehicle access to a busy town centre. Officials may examine evidence concerning air pollution, commercial activity, pedestrian safety, and public transport. The relevant propositions become significant not simply because their meanings are understood, but because they are assigned roles within the deliberation. Accepting reliable evidence of increased pollution may support the introduction of restrictions, whereas rejecting an unsupported prediction concerning commercial losses may weaken an objection to the proposal. The resulting argument depends upon how these considerations are treated and how their consequences are evaluated. Acceptance and rejection also generate rational responsibilities, since their employment exposes the positions adopted to demands for justification. The relationship between rational responsibilities and the commitments associated with the adoption of particular positions invites comparison with John Rogers Searle and Daniel Vanderveken's (1985) account of illocutionary logic, whose central concern they characterise as follows:

> A theory of illocutionary logic of the sort we are describing is essentially a theory of illocutionary commitment as determined by illocutionary force. The single most important question it must answer is this: Given that a speaker in a certain context of utterance performs a successful illocutionary act of a certain form, what other illocutions does the performance of that act commit him to?[6]

The quotation clarifies that Searle and Vanderveken's principal concern is to determine which further illocutionary acts a speaker becomes committed to through the successful performance of a particular speech act. While their account shares with the present enquiry an interest in the relationship between rational positions and commitment, the two approaches differ fundamentally in their philosophical objectives. This work does not seek to formulate a theory of illocutionary logic or to establish the commitments generated by the performance of speech acts. Rather, it examines acceptance and rejection as fundamental inferential positions through which propositions acquire rational significance independently of their actual linguistic expression. The responsibilities associated with these positions therefore arise from their employment within reasoning rather than from the illocutionary force of particular utterances. In this sense, inferential commitment concerns the justification and consequences of adopting a proposition as a reason, rather than the performance of a speech act itself. Someone who accepts a proposition as a reason must be prepared, when appropriately challenged, to explain why it deserves that role and to consider the implications arising from its use. Similarly, rejecting a claim requires distinguishing reasoned opposition from personal dislike or arbitrary refusal. These responsibilities need not imply that every proposition must be justified afresh whenever it is employed, since ordinary reasoning frequently relies upon established information and shared assumptions. Their significance lies instead in the possibility of assessing whether particular inferential positions are warranted.[7] The philosophical significance of inferential warrant has been systematically investigated by Michael Dummett (1973), whose account of meaning connects the grounds upon which a proposition may legitimately be asserted with the inferential consequences arising from its acceptance. Although Dummett's approach may be characterised as unilateral in its emphasis upon assertion and acceptance, it provides an important philosophical foundation for understanding the relationship between inferential entitlement and rational commitment. Dummett explains this relationship as follows:

> Crudely expressed, there are always two aspects of the use of a given form of sentence: the conditions under which an utterance of that sentence is appropriate, which include, in the case of an assertoric sentence, what counts as an acceptable ground for asserting it; and the consequences of an utterance of it, which comprise both what the speaker commits himself to by the utterance and the appropriate response on the part of the hearer, including, in the case of assertion, what he is entitled to infer from it if he accepts it.[8]

[6] Searle & Vanderveken (1985) Page: 6

Dummett's account thus identifies two interconnected dimensions of inferential entitlement namely, the grounds that justify the assertion of a proposition and the consequences that follow from its acceptance. But his treatment remains principally unilateral, since it does not assign rejection an independent foundational role alongside acceptance. The present study extends this philosophical perspective by maintaining that inferential warrant concerns not only the grounds for accepting propositions and the consequences of their acceptance, but also the grounds for rejecting propositions and the consequences of their rejection. Acceptance and rejection therefore constitute two distinguishable positions within rational assessment, each possessing its own justificatory requirements and inferential significance. However, the necessity of acceptance and rejection must be carefully distinguished from the stronger claim that every individual inference explicitly requires both positions. A person may reason from accepted premises without rejecting any proposition during that particular argument, while another may investigate a disputed claim without reaching a final judgement. What matters is that the wider practice of inferential reasoning possesses distinguishable positive and negative positions through which propositions can be employed and evaluated. The necessity under consideration therefore concerns the structure of rational assessment rather than an unavoidable psychological decision concerning every statement encountered. This qualification becomes especially important when examining hypothetical reasoning, where propositions appear to function inferentially without being believed.

---

[7] For a systematic investigation of the concept of warrant, see Irani (2026b), where we develop an independent philosophical discipline termed “Warrant Theory”, concerned with the inferential legitimacy of propositions within logical analysis. The central question addressed by this discipline is what entitles a proposition to function legitimately as a premise, conclusion, or intermediate step within reasoning. Warrant is understood as inferential entitlement, distinguished from truth, belief, psychological conviction, and the performance of speech acts. Within this framework, warrant-theoretic analysis is introduced as a methodical method for examining how propositions acquire inferential standing, how that standing is maintained, transferred, restricted, or lost, and how inferential positions interact through relations of dependence, compatibility, incompatibility, and exclusion. The theory adopts a bilateral framework in which acceptance and rejection represent positive and negative inferential positions, each governed by corresponding justificatory conditions and inferential consequences. On this basis, logic is characterised as the formal and normative study of the conditions under which propositions may legitimately be accepted or rejected and of the inferential transitions warranted by such legitimacy. This conception provides the broader philosophical foundation for the current project.

[8] Dummett (1973) Page: 396

Hypothetical reasoning presents a substantial challenge to the necessity of acceptance and rejection because individuals frequently draw conclusions from propositions whose truth they neither affirm nor deny. A transport planner, for example, may investigate the consequences of a major railway station becoming temporarily unavailable, despite possessing no evidence that such a disruption will occur. The planner can examine alternative routes, estimate passenger congestion, and identify possible operational difficulties without believing that the station will close. If acceptance necessarily involved believing a proposition to be true, this examination would appear to demonstrate that genuine reasoning can proceed without acceptance. The difficulty becomes more pronounced when hypothetical reasoning is understood as an essential feature of rational planning rather than an exceptional intellectual activity. Individuals regularly consider circumstances that have not occurred, evaluate possibilities they regard as unlikely, and investigate assumptions introduced solely for argumentative purposes. These activities suggest that inferential employment must be distinguished from the formation of personal beliefs. One response is to recognise provisional acceptance as a position whose scope is restricted to a particular inquiry.[9] When the transport planner adopts the possibility of closure as a starting assumption, the proposition becomes available for drawing conclusions within the hypothetical study, while no commitment to its actual truth is undertaken. Such acceptance is provisional because it can be withdrawn when the inquiry ends and does not determine the planner's settled judgement concerning future events. Nevertheless, this explanation must avoid treating every act of imagining or entertaining a proposition as acceptance. Purely contemplating the possibility of disruption does not necessarily involve reasoning from it. Provisional acceptance becomes identifiable when the proposition is assigned a definite function as a starting point from which consequences are systematically examined. Its inferential significance therefore lies in the restricted commitment to employing the assumed content according to the requirements of the argument. The distinction is substantive rather than terminological because the reasoner must recognise which conclusions depend upon the assumption, avoid treating those conclusions as established independently of it, and remain sensitive to the conditions governing its use. Hypothetical reasoning consequently demonstrates that acceptance need not constitute permanent belief, while leaving intact the question of whether adopting an inferential position is necessary for propositions to operate as premises.

The significance of provisional positions extends beyond hypothetical acceptance because reasoning may also involve the temporary rejection of propositions following an examination of their consequences. Imagine an architect considering an alternative design for a public building that would provide additional space while reducing construction costs. The design is initially introduced as a possibility, allowing its implications for accessibility, safety, and structural stability to be investigated. Subsequent calculations reveal that one proposed arrangement would obstruct an essential emergency exit. The architect may therefore reject that arrangement within the design process, although this judgement need not amount to rejecting every possible modification of the building. The rejection arises from an identified incompatibility between the proposal and an independently recognised requirement. This example illustrates an important feature of negative inferential positions since rejection is not simply a response to an undesirable consequence, but requires grounds for determining why the relevant proposition should be excluded. A hypothetical assumption may generate inconvenient implications without thus becoming irrational or unacceptable. Only when those implications conflict with considerations that possess appropriate justificatory authority does rejection acquire a defensible inferential basis. Furthermore, the provisional character of rejection permits rational reconsideration when the circumstances of an inquiry change. If a revised architectural arrangement removes the obstruction, the earlier grounds for rejecting it may no longer apply. Such revision does not show that the initial rejection was meaningless or arbitrary; it demonstrates that inferential positions can be sensitive to evidence, context, and the purposes of investigation. Acceptance and rejection should therefore be understood neither as immutable attitudes nor as unrestricted permissions to affirm and deny propositions. They establish roles whose rational legitimacy depends upon the considerations available within a particular argument. The distinction between provisional and settled positions also explains how an individual may investigate competing possibilities without endorsing their truth outside the inquiry. Yet the possibility of withdrawing either position raises a further philosophical difficulty. If rational investigation can continue while a person remains undecided about the proposition under examination, it becomes necessary to explain whether suspended judgement constitutes an alternative to acceptance and rejection or reveals a limitation in the claim that these positions are indispensable to inferential reasoning.

[9] Provisional acceptance should be distinguished from the assertion of a conditional proposition. In hypothetical reasoning, a proposition may be temporarily employed as an assumption from which consequences are investigated without being accepted as true outside the relevant inquiry. This inferential procedure does not itself require the reasoner to assert a corresponding conditional proposition. The distinction therefore concerns the inferential employment of assumptions rather than the logical properties of conditional statements.

Suspended judgement is a genuine rational possibility that must not be confused with either rejection or an inability to reason. Consider an investigator examining conflicting eyewitness accounts of a road accident. One witness maintains that a vehicle entered an intersection after the traffic signal had changed, while another insists that the signal remained favourable to the driver. If neither account is supported by sufficiently reliable evidence, the investigator may reasonably withhold judgement concerning the disputed event. This suspension does not amount to rejecting both testimonies, since the available information may provide no adequate grounds for excluding either account. Nor does it involve accepting their competing claims as simultaneously correct. The investigator instead recognises that a determinate judgement would exceed what the evidence presently justifies. Such cases appear to undermine the necessity of acceptance and rejection because the investigation may continue despite the absence of either position towards its central proposition. The objection can be addressed by distinguishing the judgement ultimately sought from the inferential positions employed during the inquiry. Although the investigator remains undecided about the traffic signal, the investigation may proceed by accepting independently verified facts, rejecting demonstrably inaccurate information, and provisionally examining the consequences of each witness's account. These activities involve distinct treatments of propositions without requiring a settled position concerning the event itself. Suspended judgement is therefore compatible with the employment of acceptance and rejection elsewhere within the same investigation. This response does not establish that every particular inference must contain both positions, nor does it justify describing uncertainty itself as a form of provisional acceptance. Rather, it clarifies the scope of the necessity claim namely, acceptance and rejection constitute fundamental kinds of inferential position available for the rational assessment of propositions, while an individual may legitimately refrain from adopting either towards a particular claim. The possibility of suspended judgement also introduces an important normative constraint. Rational reasoning must remain responsive to the limits of available evidence, since accepting or rejecting a proposition without adequate grounds may undermine the very standards of justification that make inferential positions significant. The necessity of these positions must therefore be compatible with the rational permissibility of remaining undecided.

The practical significance of inferential positions becomes particularly evident in deliberations where different considerations must be evaluated without prematurely determining the outcome. Suppose a parent is deciding whether to enrol a child in an unfamiliar educational programme. The available information may concern the instructors' qualifications, the activities offered, financial commitments, and the experiences of previous participants. Although understanding this information enables the parent to recognise the available options, reaching a rational decision requires assessing the evidential standing of particular claims. Independently verified qualifications may provide grounds for accepting that the programme meets certain educational standards, while an advertisement promising guaranteed academic improvement may be rejected because its claims exceed the available evidence. Questions concerning the child's personal interests, however, may remain unresolved until further information is obtained. The parent can consequently evaluate individual considerations without prematurely accepting or rejecting the overall recommendation. This process illustrates an important feature of practical reasoning that is to say, inferential positions may be adopted selectively, according to the evidential standing and relevance of particular propositions, rather than being applied uniformly to every claim within a deliberation. A different application arises in scientific research, where researchers examining an unexpected environmental change must distinguish observed phenomena from proposed explanations. Reliable measurements may establish a basis for further investigation, while competing causal hypotheses can be examined under specified assumptions. If additional observations undermine one explanation, researchers may have grounds for excluding it without thus establishing the correctness of its alternatives. The significance of this distinction is that rejecting one explanation does not automatically justify accepting another, even when the available hypotheses appear to exhaust the initial possibilities. Further investigation may reveal overlooked factors, methodological limitations, or previously unrecognised alternatives. Scientific inquiry therefore illustrates how the rational treatment of propositions depends upon their particular evidential functions and the circumstances under which they are employed. These examples disclose a further philosophical consideration since inferential positions are not simply classifications assigned to statements but can regulate the direction and limits of an ongoing investigation. Accepting a consideration may authorise particular lines of inquiry, rejecting another may prevent unsupported conclusions, and withholding judgement may preserve possibilities requiring further examination. The rational significance of these operations lies in their capacity to organise deliberation while remaining responsive to changing evidence, thus allowing reasoning to proceed without requiring a comprehensive or final judgement concerning every proposition involved.

To conclude, the present analysis has sought to establish why acceptance and rejection constitute two fundamental positions necessary for inferential reasoning, and why understanding propositional content alone cannot account for its employment. Its central objective has been to distinguish the comprehension of a proposition from the adoption of an inferential position towards it, without reducing that position to belief, conviction, or the performance of a speech act. The analysis has shown that acceptance enables a proposition to operate as a premise, whereas rejection establishes a negative position through which its admissibility, correctness, or consequences may be challenged. These functions are distinguishable because neither rejection nor acceptance can be identified purely with the absence of the other. Furthermore, both positions introduce rational responsibilities concerning the grounds upon which propositions are employed and the consequences arising from their employment. The examination of inferential warrant has demonstrated the importance of assessing not only the legitimacy of accepting propositions but also the legitimacy of rejecting them. This account does not require both positions to occur simultaneously within every individual inference. Rather, their necessity concerns the wider structure of rational assessment, within which propositions may acquire different inferential roles according to the evidential circumstances of an inquiry. The resulting distinction between propositional understanding and inferential employment provides the basis for addressing cases in which reasoning appears possible without a settled judgement. It also clarifies why the philosophical significance of acceptance and rejection cannot be adequately captured by an account concerned exclusively with positive commitment or assertion.

The examination of hypothetical reasoning and suspended judgement has clarified the scope of this necessity claim. Hypothetical investigation permits propositions to function as provisional assumptions without requiring belief in their actual truth, while provisional rejection allows particular possibilities to be excluded when their consequences conflict with justified requirements. Such positions remain restricted to their inquiries and may be revised when the considerations change. Suspended judgement, by contrast, remains a legitimate response where evidence does not warrant either accepting or rejecting the proposition under examination. Its permissibility does not undermine the fundamental roles of acceptance and rejection, since reasoning may continue through the selective employment of other propositions while judgement concerning the central question remains unresolved. The examples drawn from practical deliberation and scientific research have illustrated how these positions regulate inquiry without requiring premature or comprehensive conclusions. These results also indicate a number of directions for future research. One concerns the precise conditions under which provisional acceptance becomes an inferential commitment rather than mere contemplation. Another concerns the justificatory standards governing provisional rejection and the revision of negative positions. Further investigation could examine the relationship between suspended judgement and the inferential treatment of competing hypotheses. A further question concerns how the philosophical distinction between acceptance and rejection might be represented within formal bilateral systems without identifying rejection with propositional negation or reducing inferential positions to speech acts. Such investigations would extend the present account while preserving its central conclusion since propositions function within reasoning through rationally assessable inferential positions, not through understanding alone.

## References

Dummett, M. 1973. *Frege: Philosophy of Language*. London: Duckworth.

Gomolińska, A. 1998. 'On the Logic of Acceptance and Rejection', *Studia Logica: An International Journal for Symbolic Logic.* Vol. 60. pp. 233–251. https://doi.org/10.1023/A:1005009115432

Irani, K. 2025. 'Logical Inferentialism & Attacks on Classical Logic', *Open Journal of Philosophy*. Vol.15. No.4. pp. 788-813. https://doi.org/10.4236/ojpp.2025.154048

Irani, K. 2026a. 'Acceptance & Rejection', (Under Review) *arXiv:2609.26557v1*. https://doi.org/10.48550/arXiv.2609.26557

Irani, K. 2026b. 'Warrant Theory', (Under Review) *arXiv:2609.11667v1*. https://doi.org/10.48550/arXiv.2609.11667

Rumfitt, I. 2000. ''Yes' and 'No'', *Mind.* Vol. 109. No. 436. pp. 781-823. https://doi.org/10.1093/mind/109.436.781

Searle, J. R. & Vanderveken, D. 1985. *Foundations of Illocutionary Logic*. Cambridge: Cambridge University Press.